# Holder Recommendations using Graph Representation Learning & Link Prediction

Rachna Saxena
Fidelity Investments
Bangalore, India
Rachna.Saxena@fmr.com

Abhijeet Kumar
Fidelity Investments
Bangalore, India
Abhijeet.Kumar@fmr.com

Mridul Mishra
Fidelity Investments
Bangalore, India
Mridul.Mishra@fmr.com

## ABSTRACT

Lead recommendations for financial products such as funds or ETF is potentially challenging in investment space due to changing market scenarios, and difficulty in capturing financial holder's mindset and their philosophy. Current methods surface leads based on certain product categorization and attributes like returns, fees, category etc. to suggest similar product to investors which may not capture the holder's investment behavior holistically. Other reported works does subjective analysis of institutional holder's ideology. This paper proposes a comprehensive data driven framework for developing a lead recommendations system in holder's space for financial products like funds by using transactional history, asset flows and product specific attributes. The system assumes holder's interest implicitly by considering all investment transactions made and collects possible meta information to detect holder's investment profile/persona like investment anticipation and investment behavior. This paper focusses on holder recommendation component of framework which employs a bi-partite graph representation of financial holders and funds using variety of attributes and further employs GraphSage model for learning representations followed by link prediction model for ranking recommendation for future period. The performance of the proposed approach is compared with baseline model i.e., content-based filtering approach on metric hits at Top-k (50, 100, 200) recommendations. We found that the proposed graph ML solution outperform baseline by absolute 42%, 22% and 14% with a look ahead bias and by absolute 18%, 19% and 18% on completely unseen holders in terms of hit rate for top-k recommendations: 50, 100 and 200 respectively.

## 1 INTRODUCTION

Recommender systems (RS) for products had been successfully integrated into e-commerce like movie or retail product recommendations. Numerous RS methods had been proposed including the popular techniques: Content based filtering, Collaborative filtering (item based, user based), Model based filtering with deep learning, knowledge graph-based RS [1][2][3][4]. Content based filtering has difficulty in generating diversified recommendations (only recommends similar characteristics products) while collaborative filtering only recommends products bought by similar users and fails to recommend indirect products. Most of the real-world systems are combination of two techniques to leverage advantage of both, hence ensuring the recommender is smarter [5].

Recent advancement in Graph Machine Learning (GML) techniques have enabled researchers to apply new ways to solve problems in financial domain like fraud detection, fund networks, community detection etc. These are variety of problems which can be solved using supervised, semi supervised, or unsupervised techniques based on underlying knowledge graph construction. With advent of Graph Neural Networks (GNN), researchers have been exploring representation learning methods to incorporate both content attributes (characteristics) similarity and structural (behavioral interest) similarity in a single model [6]. GraphSAGE (Hamilton et al, NIPS 2017) [7] is a representation learning technique for which uses inductive learning based on the local features and neighborhood of the node. It is popular research method for link prediction task where node local features (content attributes) are of significance. This paper explores and demonstrate an application of GML based representation learning and link prediction model in financial use-case of holder lead recommendation system.

Why does a holder (private banks, fund manager, family offices, registered investment holders, advisory firms) invests in a particular investment product? There might be multitude of factors influencing the investment decision ranging from industry anticipation, product categories, investment style, macro factors, existing relationships with product issuers etc. [8][9] Moreover, it also depends on asset flows or capital available to invest from customers or retail investors [10]. Finally, with multiple competitive products in market, a holder may invest in apt product based on specific attributes like fees, risk, returns. We propose a data driven comprehensive framework to generate holder leads. High level components ad presented in Fig. 1.

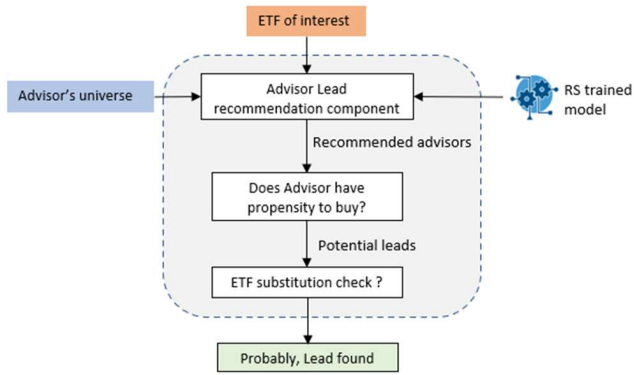

Figure 1: Holder Lead Generation Framework for Product

This paper focusses on implementation details and performance evaluation of holder lead recommendation component. The component aims to generate lead recommendations (holders) given a fund. Bi-partite graphs with node features is a useful data structure to represent such scenarios as shown in Fig. 2. Holder profile similarity can be modelled using node attributes and investment interest can be represented by edges (based on the investment made).

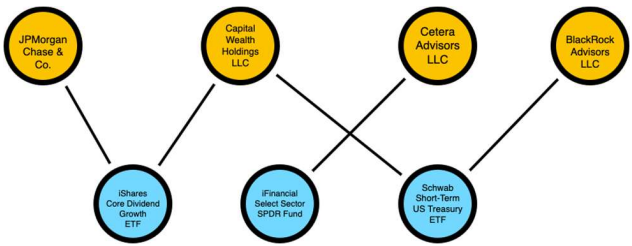

Figure 2: Holders-ETF bipartite graph

## 2 RELATED WORKS

The adoption of graph machine learning techniques like GCN, GraphSAGE [6][7] in financial applications has been less explored and requires traction in literature. There has been extensive research work in the field of graph machine learning models in recent years. For structural similarity, variants of translation models were proposed in past e.g., TransE [11], TransH [12], and TransR [13]. Further, evolution of factorization-based or multiplicative models took place which performs low rank factorization to create node embeddings like DistMult [14], RotatE [15] and ComplEx [16]. All the mentioned models do not account for representation of node features/attributes while learning representation. Heterogeneous construct where features are modelled as nodes fails to show results on downstream tasks. With advent of GCN and its variant, there had been promising results [6][7] as these techniques learn representation incorporating local features as well as structural similarity using neighborhood.

To our knowledge, there has been few existing research on GML models for recommendations systems for user-item interactions. Yue Deng did comprehensive review of recommendation systems using graph embedding techniques [1]. The paper systematically retrospect graph embedding-based recommendation from embedding techniques for bipartite graphs, general graphs and knowledge graphs and concluded that conventional models can still outperform graph embedding ones on user item interaction tasks. Muhammad Umer et al. trained multi-layer GNNs with meta-path aware information and shows outperformance on several user-item interaction datasets [2]. Mahdi Kherad et al. applied autoencoders and deep learning methods on user trust graph (user-item interactions) [3]. Sai Mitheran et al. proposed a technique that leverages a Transformer in combination with a target attentive GNN and showed competitive results with existing methods on real world benchmark datasets [4].

Moreover, there had been various research to apply GML in finance domain. Vipul Satone et al. applied node2vec (structural similarity) to perform fund level similarity based on underlying securities [17]. On similar lines, Bhaskarjit Sarmah et al. employed node2vec for stock correlation using its log returns correlation graph for SNP500 [18]. Olakunle Temitope et al. proposed a KNN based recommender framework for the banking domain that recommends customized products from the experience and historical transactions of the observed customers stored in graph-oriented database [19][20]. Hongwei Wang et al. employed variant of PageRank algorithm in combination with a bipartite graph-based collaborative filtering to develop recommender for crowdfunding campaign (user-campaign interaction) [21]. Jingming Xue et al. proposes a group recommendation model based on financial social networks and collaborative filtering algorithms to build a robo-advisor [22][23][24].

Lastly, there are studies done by finance academic researchers to understand the financial advisor's investment philosophy. Patrick Bolton et al. performed extensive study of institutional investor's ideology in left-right dimensions where left supports more social and environment-friendly orientation of the firm and money conscious investors appears on the right [9]. Leonard Kostovetsky et al. found that passive ETFs from large index providers (brand names only) attract more capital and strong preference from institutional investors and no preference from retail investors in their work [10]. Jonathan Brogaard et al. discovered that model portfolio provided by ETF issuers has substantial impact on ETF flows [8]. However, conflicts of interest (as asset managers include their own affiliated ETFs) seem to affect the quality of these recommendations. The domain studies regarding holder-funds interactions and research utilizing GML models in recommendations discussed in this section motivated the novel work on holder lead generation system presented in the paper. The outline of the rest of paper is as follows: The next section describes dataset, graph construct and proposed graph machine learning based recommender in detail. Section 4 shows the evaluation setup, results, comparative study of model variants and its analysis.

## 3. PROPOSED METHODOLGY

### 3.1 Problem Definition & Graph Construction

In this recommendation scenario, we have a set of holders: A = {a1, a2, a3 … aM} and set of funds: F = {f1, f2, f3…fN}. Our knowledge graph G is an undirected graph G = (V, E) where V ∈ (A ∪ F) and E = {e1, e2 …eK} is the set of edges. We define edge between two nodes as e = $(u, v)$ such that $u$ ∈ A, $v$ ∈ F and $(u, v)$ ∈ E. Edge between nodes $u$ and $v$ means there exist a path between them in the knowledge graph. We also assign true label to edges: y(e) ∈ {0, 1} where y(e)=1 means, there exists a relation between holder and fund entities in the observed data. Contrarily y(e)=0, means corresponding holder and fund entities are not related to each other in the observed data.

Using bipartite graph design, we define our problem as, predict the probability of having an edge between a pair of nodes where source node belongs to holder set and destination node belongs to fund set.

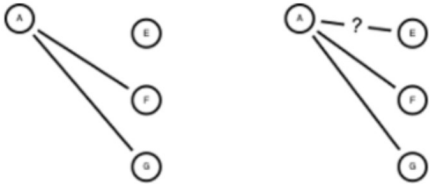

Figure 3: Training at time T and Prediction at time T+△t

GML techniques consists of algorithms which uses graph property of data to solve different problems. By graph property we mean, presence of entities and relationship between those entities. Within GML domain, GNN provide set of algorithms which learn from local and neighborhood features of the entities by message passing mechanism and generate representation for different graph entities [25].

To solve our problem of predicting link between different entities of a graph, we use GraphSAGE [26] variant of GNN. In this algorithm, aggregator functions use features of neighboring nodes to learn distribution of node features. This neighborhood-based training enables aggregate functions to generate node embeddings of unseen nodes during inference. For the downstream task of predicting link between nodes, we train a multi-layer perceptron (MLP) model [27]. We create unidirectional, bipartite graph using fund and holder entities (Fig. 2) as nodes and normalized data features as node attributes as discussed in section 3.2.

### 3.2 Dataset & Feature Engineering

For evaluation purpose, we collected holder's transactions reported data from quarterly filed 13F forms. 13F filing are mandated by SEC for all financial holders to disclose their investments having more than $100 million assets under management. We gathered 5K holders invested in 2K+ unique products along with market value invested in products, product category, investment strategy (active, passive, strategic), product issuer for 5 years.

Since all attributes were categorical, features were transformed with one hot encoding and actual market values were uses instead of binary encoding. Further, we aggregated the attribute representation at holder level as well as at fund level. In total, there were 350+ features on which we performed min max scaling to bring invested amount on same scale. Exactly same features were employed for both content filtering baseline model as well as proposed representation learning approach (as node attributes for holder & fund nodes, refer Fig. 2). Holder AUM segments were also created for diverse recommendations.

### 3.3 Learning Model

This section discusses the implementation and model learning details of our proposed approach. The overall training process constitute graph construction, node feature engineering, combined training of GraphSage & link prediction model. Further, inferencing constitutes of generating link confidence scores and ranking the recommendation in descending order. The overall flow has been presented in Fig. 4.

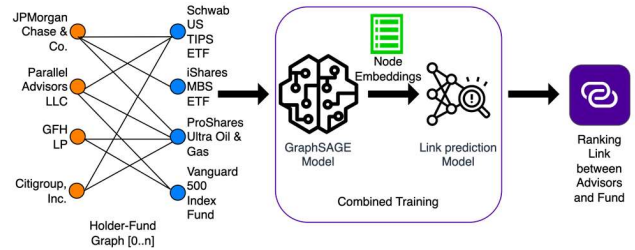

Figure 4: Graph Representation learning & Link Prediction Model

We implement GraphSage model based on Deep Graph Library [28] (DGL) using Pytorch framework [29]. Our model consists of multiple SAGEConv layers. DGL implements SAGEConv using GraphSAGE [7] algorithm. Mathematical formulation of same can be seen in the equation below.

$$h_v^k = \sigma[W^k \cdot CONCAT(h_v^{k-1}, AGGREGATE_k(h_u^{k-1}, \forall u \in \mathcal{N}(v))]$$

Given input undirected graph and node features; this algorithm learns node embeddings without node labels. Based on input configuration, different aggregate functions are learnt during training. These functions implement diverse ways of incorporating neighborhood node feature information in the learning process. We experimented with all available aggregated functions like mean, pool, GCN and LSTM [30].

Our GraphSage model takes graph input and generate node embedding output. For validating node embeddings, we create positive graph from the valid edges of original graph. Similarly, negative graph is created from non-existing edges of the original graph. At each training step, node embeddings output from GraphSage model are assigned to both positive and negative graphs.

Together with GraphSage, we also train a task specific MLP model called Link Prediction (LP) model. As name suggests, it performs the downstream task of predicting link between nodes. First, it calculates edge embeddings by concatenating pair of node embeddings and then learns to assign a scaler score to each edge. This score represents the probability of edge being present between pair of nodes as shown in equation below.

$$y_{u \sim v}^{pred} = [W_2 . \max(0, W_1 . CONCAT(u, v))]$$

We use predicted edge scores and true edge labels to calculate combined binary cross-entropy loss for both the models. Adam optimizer [31] is used to update weights during back propagation. The loss can be represented by the below equation.

$$L_{BCE} = -1/n \sum_{i=1}^{n} (y_i . \log(y_i^{pred}) + (1 - y_i) . \log(1 - y_i^{pred}))$$

During training, learning rate remains constant at 0.01 and output node embeddings dimension is set to 128. We get AUC of 69% on our test set of positive and negative graphs. The computed loss is plotted in the Fig. 5 below.

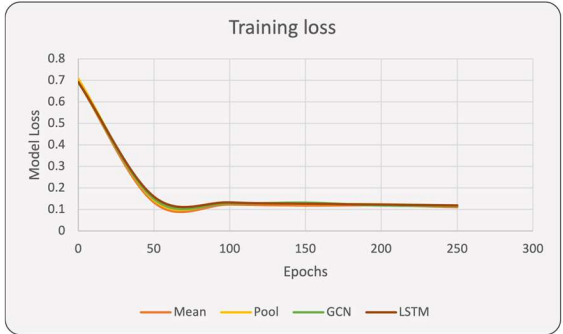

Figure 5: Training loss for different aggregate functions

## 4. EXPERIMENTS & RESULTS

The comparative experiments between content filtering-based baseline approach and proposed GML approach were done exhaustively for all available ETFs to establish the superior performance of latter on 4 qtrs. of data. Training was done in point of time (specific qtr.) and we evaluated the performance by predicting Top-K holders as recommendations. The actual holders of next quarter for input funds were considered as ground truth. The ground truth comprises of existing holders (look ahead bias) from previous quarter and newly added holders.

### 4.1 Baseline Model

The baseline system is a cosine-based content attribute similarity model to generate lead recommendations based on holder investment profiles. The features employed and transformations are presented in Section 3.2. The baseline model is more transparent but lacks in accurate lead generation and does not use Holder Fund preference directly (as opposed to collaborative filtering approach). We created a variant of baseline where holder segments constraint was imposed on results meaning recommendation were provided in same proportion of holder AUM segments as in training data (previous qtr.). Fig. 6 shows an intuitive design of baseline model for cosine similarity search. The markers show investment made by holders in product characteristics and highlighted rows show holder similarity in their vector representations.

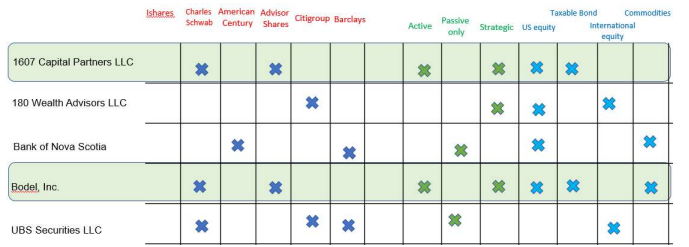

Figure 6: Abstract vector space baseline model

### 4.2 Results & Analysis

Hits in Top-K (K: 50, 100, 200) were used for performance evaluation. The hit rates are defined as percentage of recommendations present in ground truth divided by minimum (K, # holders invested in fund). The hit rates are presented in Table I. One can observe that GraphSage implementation outperforms baseline by absolute 42%, 22% and 14% in terms of hit rate for top-k recommendations: 50, 100 and 200 respectively. The metrics achieved for top performing GraphSAGE models (GCN, LSTM) were consistent for top k (50, 100, 200). Though as an established fact, GCN aggregator training computationally takes significantly less time than LSTM aggregators in terms of time efficiency.

TABLE I. EVALUATION ON 13F DATASET (2021Q4)

| Approach | Performance metric Hits @ Top K | | |
|---|---|---|---|
| | Next Qtr. *Hits@top50* | Next Qtr. *Hits@top100* | Next Qtr. *Hits@top200* |
| Baseline Model | 21.6% | 25% | 32.8% |
| **Baseline Model (diversity constraint on holder segment)** | **34%** | **43%** | **53.5%** |
| GraphSage (mean) + MLP | 19% | 25% | 33% |
| GraphSage (pool) + MLP | 57% | 62% | 67.5% |
| **GraphSage (GCN) + MLP** | **63%** | **65%** | **67.7%** |
| GraphSage (LSTM) + MLP | 62% | 65% | 67.6% |

There were two major takeaways from the experiments. First, training the GraphSage parameters and link prediction model parameters separately fails to recommend better leads in comparison to baseline. Training weights for GraphSage tied with link prediction parameters optimizes the loss function properly (Fig. 4), hence generating better leads. Second, out of various aggregators (parameter for sample neighborhood aggregation), GCN and LSTMs performed better than mean aggregator or pooling [30]. Mean aggregator and dot predictors (link prediction) did not work in this use-case.

In another set of experiments, we established through results in Table II that model did learn new information i.e., likelihood of a non-existing holder to invest in financial product in future time. Motivation behind the experiments was the fact that a portion of generated lead were already existing holders of the product hence capturing a look ahead bias. Table II shows hits in Top-K for news added holders which model has not seen while training. Interestingly, the proposed graph-based approach outperforms the content filtering approach by absolute 18-19% hit rate.

TABLE II. EVALUATION ON NEWLY ADDED HOLDER 13F DATASET

| Approach | Performance metric Hits @ Top K | | |
|---|---|---|---|
| | Newly added Next Qtr. *Hits@top50* | Newly added Next Qtr. *Hits@top100* | Newly added Next Qtr. *Hits@top200* |
| Baseline Model (diversity constraint on holder segment) | 8.8% | 15.5% | 24.3% |
| GraphSage (GCN) + MLP | 27% | 35% | 43% |

The recommendations (holders) surfaced by the recommendation component developed can be considered as potential leads only. There are various other factors which would play a pragmatic role in conversion rate (product sales) as discussed in Section 1. For example, the macro-economic factors supporting the product sales, propensity of a holder to buy the product (asset flows). From reasoning perspective, system surface the alternative products in holder's portfolio in case they are not existing holders of queried product. Further, proposed system checks the fees and risk-return rewards and finally generate prospective sales leads if the metrics are better than alternative product.

## 5. CONCLUSION & FUTURE WORKS

The paper explores an uncharted area of applying advance graph machine learning for lead generation task in finance. The penned work is part of comprehensive data driven framework for developing a holder recommendations system for financial products like ETFs or funds. This paper proposes and implements an overall bipartite graph-based learning model which incorporates node feature engineering, representation learning and link prediction task for a financial holder lead recommendation system. The presented work re-established that combined training of representation learning and link prediction models for supervised task leads to appropriate estimation of model weights and the same is not learnt properly when models are separately trained sequentially. We demonstrated through experiments that learnable aggregators like GCN and LSTM performs better than well-defined functions like mean and pooling aggregators in the prediction performance. We conducted exhaustive experiments to evaluate the holder lead generation for fund universe unseen newly added holder for respective funds for next quarter. Results exhibits signification improvement over content filtering-based baselines models by absolute 18-19% hit rate in top-K recommendations.

As future directions, we would like to extend the work further on incorporating the timeliness factor (period) as current work learns at a point in time. Moreover, as GraphSAGE is an inductive framework [7], we intend to evaluate results of implemented recommender on new product launches. One can improve upon the feature engineering process to create new useful holder attributes to model their investment philosophy like demographic, education, social relations, professional backgrounds, or fund attributes etc. On the GML front, we look forward to incorporate edge attributes to model strength of preferred products by holder in bi-partite construct and learning model to accommodate unequal holder and product node attributes, exploring heterogenous graph properties etc.

## ACKNOWLEDGMENTS
We used Deep Graph Library (DGL) extensively for implementing the proposed solution [28]. Our sincere gratitude to the owners and all the contributors of DGL for the magnificent work.